\documentclass[twocolumn]{revtex4}
\usepackage{graphicx}
\usepackage{color}
\usepackage{amssymb,amsfonts,amsmath}

\begin{document}

\title{Congruent evolution of genetic and environmental robustness in microRNA}

\author{Gergely J.\ Sz\"oll\H{o}si}
\email{ssolo@angel.elte.hu}
 \affiliation{E\"otv\"os University, Budapest,
Hungary P\'azm\'any P\'eter S\'et\'any 1/A.}
 \homepage{http://angel.elte.hu/~ssolo}
\author{Imre Der\'enyi}%
 \email{derenyi@angel.elte.hu}
\affiliation{E\"otv\"os University, Budapest, Hungary P\'azm\'any P\'eter S\'et\'any 1/A}
 \homepage{http://angel.elte.hu/~derenyi}

\date{\today}

\begin{abstract} Genetic robustness, the  preservation of  an optimal
phenotype in the  face of mutations, is critical  to the understanding
of evolution as phenotypically expressed genetic variation is the fuel
of  natural selection. The  origin of  genetic robustness,  whether it
evolves directly by natural selection  or it is a correlated byproduct
of  other  phenotypic  traits,  is,  however,  unresolved.   Examining
microRNA (miRNA)  genes of several eukaryotic  species, Borenstein and
Ruppin (Borenstein et al.\ 2006, PNAS 103:\ 6593), showed that the
structure  of miRNA precursor stem-loops exhibits  significantly
increased mutational  robustness in comparison  with  a sample  of
random  RNA  sequences with  the  same stem-loop  structure.   The
observed   robustness  was  found  to  be uncorrelated with
traditional  measures of environmental robustness -- implying that
miRNA sequences show evidence of the direct evolution of genetic
robustness.   These findings  are  surprising as  theoretical results
indicate that the direct evolution of robustness requires high
mutation  rates and/or  large  effective population  sizes only  found
among RNA viruses, not  multicellular eukaryotes.  We demonstrate that
the  sampling   method  used  by  Borenstein   and  Ruppin  introduced
significant  bias  that  lead  to  an  overestimation  of  robustness.
{\color{black} Introducing a  novel measure of environmental robustness
based on the equilibrium  thermodynamic  ensemble of  secondary
structures of  the miRNA precursor sequences we demonstrate that the
biophysics of RNA  folding,  induces a high level of correlation
between genetic (mutational)  and environmental (thermodynamic)
robustness, as expected from the theory of plastogenetic congruence
introduced by Ancel and Fontana (Ancel et al.\ 2000, J.\ Exp.\ Zool.\
288:\ 242).}  In light of  theoretical considerations we believe  that
this correlation strongly suggests that genetic robustness observed in
miRNA sequences is the byproduct of selection for environmental
robustness.
\end{abstract}


\maketitle

\section*{Introduction}

The magnitude of genetic effects on phenotype depends strongly on genetic background, the effects of the same mutation can be larger in one genetic background and smaller in another. The idea that wild-type genotypes are mutationally robust, i.e.\ show invariance in the face of mutations (more generally heritable perturbations), goes back to Waddington \cite{waddington}, who originally introduced the concept as canalization. While genetic robustness has been found across different levels of organization from individual genes, through simple genetic circuits to entire organisms (approximately $80\%$ of yeast single knockouts have no obvious effect in rich medium \cite{hillenmeyer}), the origin of the observed robustness has remained a source of contention. The three main hypotheses regarding the potential origin of genetic robustness predate the concept itself, and fall along the lines of the famous debate between members of the modern synthesis (in particular Wright, Haldane and Fisher) surrounding the origin of dominance (dominance can be understood as a simple case of robustness, a dominant phenotype being more robust against mutations) \cite{devisser,burger}: (i) the most straightforward explanation, favoured by Wright, was that robustness evolves \emph{directly}, through natural selection \cite{fisher}; (ii) an alternative \emph{congruent} hypotheses, put forward in the context of dominance by Haldane, proposes that the evolution of genetic robustness is a correlated byproduct of selection for environmental robustness, i.e.\  invariance in the face of nonheritable perturbations, e.g.\ temperature, salinity or internal factors such as fluctuations in the concentration of gene products during development \cite{haldane}; (iii) while a third view holds that  genetic robustness is \emph{intrinsic}, arising simply because the buffering of a character with respect to mutations is the necessary or likely consequence of character adaptation, in the context of dominance Wright \cite{wright}  and later Kacser and Burns \cite{kacserburns} argued that it arises as an inevitable, passive consequence of enzyme biochemistry and selection for increased metabolic flux. 

Recently robustness has been a subject of renewed interest.
Several theoretical and simulation studies have addressed robustness in a wide range of contexts ranging from gene redundancy \cite{krakauerPNAS} to model regulatory networks \cite{siegalPNAS,azevedoNAT,wagnerPLOS,wagnerPNAS,crombachPLOS}. {By in the first case building on evidence of excess mutational robustness present in RNA secondary structure \cite{wagnerJexp} and in the second case on the expectation that high mutation rates present among RNA viruses should favour mutational robustness \cite{wilkeadami}} two pioneering studies, by Montville et al.\ \cite{montville} and Borenstein and Ruppin \cite{ruppin} have managed to step beyond computer simulations and through using, respectively,  \emph{in vitro} evolution experiments \cite{montville} and microRNA sequences from diverse taxa found evidence to support the hypotheses that genetic robustness can evolve directly. {\color{black}Further work on \emph{in vitro} evolution experiments has provided additional evidence showing that if a population is highly polymorphic robustness can evolve directly \cite{sanjuan,bloom}. }   

The theoretical underpinnings of these studies is provided by the results of van Nimwegen et al.\ \cite{nimwegen}, who through solving the quasispecies equations describing the evolution of a population on a network of phenotypically neutral sequences, were able to demonstrate, that provided a sufficiently polymorphic population, mutational robustness can evolve directly. The necessary mutation rates and/or population sizes were found to be very large in simulation studies using RNA secondary structure as a genotype-phenotype map \cite{nimwegen,adamiJTB,szollosiMBS}, direct evolution of increased neutrality requiring the product of the effective population size $N_e $ and the mutation rate per nucleotide $u$ to be well in excess of one. Such high mutation rates can only readily be found among RNA viruses, are extraordinary even among unicellular organisms (\emph{Prochlorococcus}  $2N_e u\approx2.$, \emph{E.\ coli} $2N_e u\approx0.2$, \emph{S.\ cerevisiae}  $4N_e  u\approx0.09$) and completely unheard of among multicellular eukaryotes possessing RNA silencing mechanisms and microRNA genes (\emph{A.\ thaliana} $4N_e u\approx0.012$, \emph{D.\ melanogaster} $4N_e u\approx0.015$, \emph{C.\ elegans} $4N_e u\approx0.013$, \emph{C.\ intestinalis} $4N_e u\approx0.012$, \emph{M.\ musculus} $4N_e u\approx0.001$, \emph{H.\ sapiens} $4N_e u\approx0.001$) \cite{lynch}.    

In their study Borenstein and Ruppin examined microRNA (miRNA) precursor sequences from several eukaryotic species. miRNA are small endogenous noncoding RNAs that regulate the expression of protein coding genes through the RNA interference (RNAi) pathway \cite{lagosQSCI,lauSCI,leeSCI,bartelCELL}. Functionally relevant short ($\approx22$nt) mature miRNA sequences are excised from longer precursor sequences that fold into a stem-loop hairpin structure. The hairpin like secondary structure of precursor stem-loops plays a crucial role in the maturation process \cite{bartelCELL} and is under evolutionary constraint to conserve its structure. Borenstein and Ruppin used the novel and ingenuous method of generating for each miRNA sequence a random sample of sequences with identical minimum free-energy (MFE) structure to uncover traces of adaptation. To compare the mutational robustness of miRNA precursor sequences to random sample sequences with identical MFE structure they compared the single mutant neighborhood of a given miRNA precursor sequence to the single mutant neighborhood of the sample sequences. Calculating the average distance of the MFE structure of each single mutant sequence to the MFE structure of the original sequence for both stem-loop and sample sequences {\color{black} (details on secondary structure calculations are presented below)} they demonstrated that miRNA precursor sequences have single mutant neighborhoods with sequences that fold into more similar MFE structures compared to sequences in the single mutant neighborhoods of sample sequences with identical MFE structure. While a similar comparison of the folding minimum free-energy showed a comparable, but lower bias, the finding that the two were only weakly correlated allowed the authors to conclude that the observed bias is a result of direct selection for mutational robustness.  {\color{black} Their results were reexamined by Shu et al.\ \cite{Shu} who argued that mutational robustness among miRNA precursors may be the correlated byproduct of selection for environmental robustness, but found only a moderately higher correlation using a different measure of mutational robustness. }

In light of the consistently low value of $uN_e$ among multicellular eukaryotes the results of Borenstein and Ruppin are highly surprising. There is no known mechanism which can explain the direct evolution of robustness that they observe. According to the classic results of Kimura and Maruyama the average fitness of an asexually reproducing population (in the limit of very large populations) depends only on the mutation rate and is independent of the details of the fitness landscape \cite{kimura}. This result, however, only holds under the assumption that the fittest genotype does not have any neutral sites. While, the extension of these results to more general fitness landscapes by van Nimwegen et al.\ demonstrates that the presence of neutral genotypes can lead to selective pressure to evolve mutational robustness simulation studies using genotype-phenotype maps induced by RNA secondary structure have demonstrated that $uN_e>1$ is a necessary condition \cite{adamiJTB} even in the presence of recombination \cite{szollosiMBS}. The case for the direct evolution of genetic robustness rests on the parallel findings that a stronger bias for mutational robustness is present in miRNA precursor sequences than for environmental robustness and that the two are only weakly correlated. Introducing a new measure of environmental robustness in this paper we endeavor to demonstrate that, indeed  as previously also suggested by Shu et al.\ \cite{Shu}, the exact opposite is true: the bias for environmental robustness is stronger and it is highly correlated with mutational robustness. {\color{black} The correlated evolution of environmental and mutational robustness in RNA sequences under selection to retain secondary} {\color{black}structure is expected as a corollary of a general casual link between environmental robustness and genetic robustness} {\color{black}in RNA sequences proposed by  Ancel and Fontana \cite{ancelfontana}. They argue that \emph{plastogenetic congruence}, i.e.\ the correlation between the set of structures thermally accessible to a sequence, its \emph{plastic repertoire}, and the MFE structures of its \emph{genetic neighbourhood} will lead to the emergence of mutational robustness in the presence of selection for some predefined structure. }

\begin{figure}
\centerline{\includegraphics[width=0.5\textwidth]{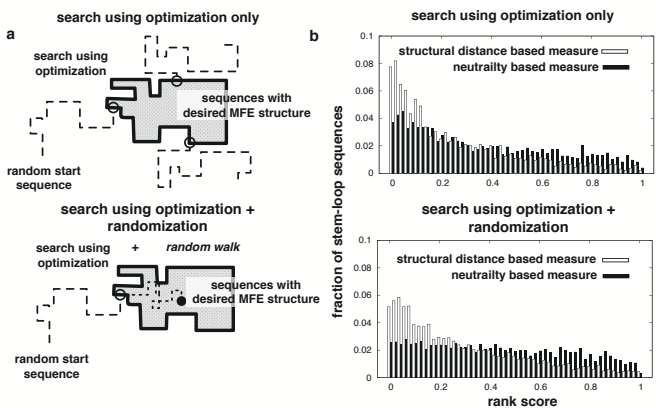}}
\caption{{\bf a} Generating a random sample of sequences with a desired MFE structure by stochastic minimization of the free-energy of the desired fold (the method employed the RNAinverse program used by Borenstein and Ruppin \cite{ruppin} as well as Shu et al.\ \cite{Shu} ) results in a biased sample in which sequences with lower than average neutrality (higher than average number single mutant neighbors) are overrepresented. This can be avoided if after finding a sequence with the desired MFE structure a random walk is performed among sequences with the desired MFE structure. {\color{black} This random walk on the neutral network associated with the MFE structure mimics the sequnce drift of a sequence evolving under the constraint to fold in to the desired MFE structure.}
{\bf b} Rank score distributions for two measures of mutational robustness ($\eta_s$ and $\eta_n$ see text). Comparing the distributions derived from sampling using only stochastic optimization (top, $\bar r^{\rm biased}_s=0.25,R^{\rm biased}_s=0.83,\bar r^{\rm biased}_n=0.37,R^{\rm biased}_n=0.66$) to that derived from sampling with subsequent randomization (bottom,  $\bar r_s=0.29,R_s=0.78,\bar r_n=0.44,R_n=0.59$) shows that increased neutrality is predominately an artifact of biased sampling, while the lower than average distance of MFE structures in the mutational neighborhood to the wild-type MFE structure becomes somewhat less pronounced, but is still significant. }\label{fig1}
\end{figure}

\section*{Materials and Methods}
\subsection*{microRNA sequences and sampling } 
 miRNA precursor sequences were downloaded from miRBase version $9.0$ \cite{mirbase}.
All $4361$ miRNA genes were used, yielding $3641$ unique miRNA precursor sequences. For each miRNA precursor sequence we produced a sample of random sequences by (i) using the stochastic optimization routine from the Vienna RNA package \cite{vienna} to produce a sequence with MFE structure identical to that of the native sequence that is stored (ii) and subsequently randomizing this sequence by attempting $4L$ random nucleotide substitutions in a miRNA precursor sequence of length $L$, accepting a substitution if the resulting sequence's MFE structure remains unchanged. For each miRNA precursor sequence on average $>800$ sample sequences with identical MFE structure was generated. Supplemental material accompanying our paper contains the robustness values for all $4361$ genes associated with $3641$ unique sequences we considered.   

\subsection*{Measuring thermodynamic robustness} 
In order to calculate {\color{black} the thermodynamic robustness measure $\eta_t$, defined below,} we sampled the equilibrium thermodynamic ensemble of stem-loop and sample sequences using the stochastic backtracking routine from the Vienna RNA package producing $10^6$ suboptimal structures per sequence, using the default temperature of $310\ K$. The average distance from the MFE structure in the thermodynamic ensemble can be calculated exactly with the help of base-pairing probabilities, which are available as a byproduct of partition function folding in the Vienna package, and were used to validate the sampling.    

\subsection*{Statistics} 
Given a rank score $r$ and sample size $N$ a good estimate for the probability of observing an equal or lower rank score by chance is given by $(r N)/(N+1)\approx r $. Following \cite{ruppin} rank scores of $r<0.05$ are considered significantly robust. To determine if the robustness of miRNA precursor sequences according to some measure $\eta$ has the same distribution as the robustness of sample sequences $\eta'$ for a group of sequences, following \cite{ruppin} we test against the null hypothesis that they are drawn from identical distributions using the nonparametric Wilcoxon signed rank test. In contrast to \cite{ruppin}, however, we do not consider as paired values $\eta$ and the average of $\eta'$ over all $N$ sample sequences, $\langle \eta' \rangle$, as we found this to result in spuriously low $p$-values, but instead calculate the $p$-values for a given group of sequences by averaging over $1000$ different sets of $\{ \eta,\eta' \}$ pairs where in each set the $\eta'$ values belong to a random sample sequence. As a complementary  approach we also tested the hypothesis that the distribution of rank scores of a group of sequences for a given robustness measure is uniform -- as we would expect if miRNA precursor sequences were randomly sampled from the set of sequences with identical MFE structure -- using a standard Kolmogorov-Smirnov goodness of fit test. We found the two significance analyses to be in good agreement indicating highly significant bias for higher values of $\eta_t(d_{\rm th.})$ and $\eta_s$, but mostly no or only nonsignificant bias for higher $\eta_n$. The supplemental information accompanying our paper contains species level statistics and significance analyses.   
           
\begin{figure}
\begin{center}
\centerline{\includegraphics[width=0.5\textwidth]{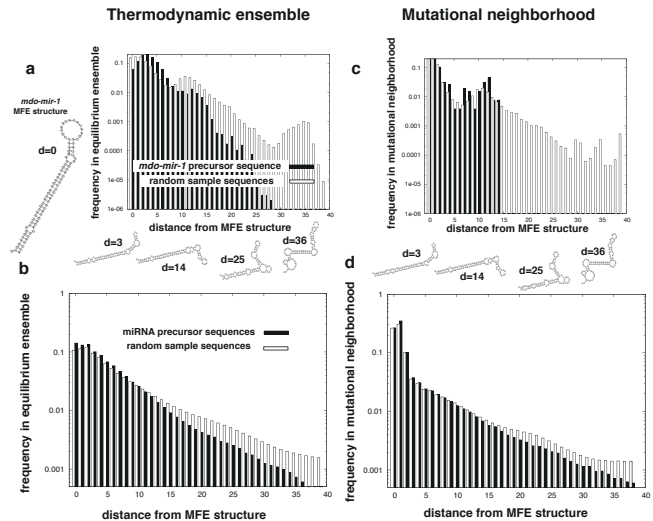}}
\caption{In order to examin the robustness of miRNA precursor sequences to thermal fluctuations we sampled the equilibrium thermodynamic ensemble of structures. Sampling $10^6$ structures for each miRNA precursor sequence and each member of the random sequence sample, we binned structures according to their distance from the MFE structure. {\bf a} For, e.g. the \emph{Monodelphis domestica} miRNA precursor sequence \emph{mdo-mir-1} examining the distribution of structures as a function of the base-pair distance shows that the averaged random sequence sample distribution (white bars) has a much larger fraction of structures that are drastically different from the MFE structure, compared to the distribution of structures for the original miRNA precursor sequence (black bars). {\bf b} Examining the averaged distribution of stem-loop (black bars) and random corresponding random sequence sample distributions (white bars) shows that there is a general tendency among miRNA precursor sequences for increased thermodynamic robustness, i.e.\ of avoiding structures that are highly dissimilar to the MFE structure. A strikingly similar effect can be observed if we examine the distribution of structures in the mutational neighborhood. Analogous to {\bf a}, in {\bf c} we binned, according to their distance from the MFE structure of the wild type, the MFE structures of all ($3L$) single point mutants for the \emph{Monodelphis domestica} miRNA precursor sequence \emph{mdo-mir-1} (black bars) as well as the MFE structures for each sequence in the single mutant neighborhood of sample sequences (white bars). The distribution of structures in both the thermodynamic ensemble ({\bf a}) and the mutational neighborhood ({\bf c}) of the \emph{mdo-mir-1} miRNA precursor sequence have a significantly smaller fraction of structures that are highly dissimilar then sample sequences with identical MFE structure. Comparing the the averaged distribution of stem-loop (black bars) and random corresponding random sequence sample distributions (white bars) in the mutational neighborhood ({\bf d}) to similar averaged distributions in the thermodynamic ensembles of the same sequences ({\bf b}) shows that the tendency among miRNA precursor sequences for increased robustness is present both in the mutational neighborhood and the thermodynamic ensemble, i.e. miRNA precursor sequences show excess robustness in the face of both thermal and mutational perturbation.}\label{fig2}
\end{center}
\end{figure}

\begin{table*}
\caption{ Phylogenetic breakdown of different measures of robustness }
\begin{tabular}{@{\vrule height 10.5pt depth4pt  width0pt}l|ccc|ccc|ccc|cc}
group / species &$\bar r_n$ & $R_n$ & $S_n$ & $\bar r_s$ & $R_s$ &  $S_s$ & $\bar r_t(25)$ &  $R_t(25)$ & $S_t(25)$ & $c(r_s,r_t(25))$ & \# of seqs.  \\
\hline\hline
all 
& {\bf 0.44}& 0.59 & 0.08 &  {\bf 0.29}& 0.78 & 0.17&{\bf 0.31}& 0.74 & 0.28 & 0.73 &  3641 \\
\hline
vertebrate 
&       0.46 & 0.55 & 0.06 &{\bf 0.31}& 0.78 & 0.13&{\bf 0.29}& 0.75 & 0.30 & 0.76 &  2215 \\
invertebrate 
&{\bf  0.37}& 0.68 & 0.10 &{\bf 0.21}& 0.88 & 0.27&{\bf 0.22}& 0.84 & 0.36 & 0.73 &  488 \\
landplant 
&{\bf  0.41}& 0.63 & 0.11 &{\bf 0.31}& 0.75 & 0.21&{\bf 0.40}& 0.63 & 0.19 & 0.68 &  848 \\
virus 
&       0.38 & 0.66 & 0.09 &{\bf 0.23}& 0.84 & 0.18&{\bf 0.21}& 0.85 & 0.32 & 0.65 &  82 \\
\hline
\emph{Homo sapiens} 
&       0.48& 0.53 & 0.04 &{\bf 0.32}& 0.75 & 0.11&{\bf 0.28}& 0.76 & 0.33 & 0.74 &  471 \\
\emph{Mus musculus} 
&       0.46& 0.56 & 0.06 &{\bf 0.33}& 0.76 &  0.12&{\bf 0.31}& 0.74 & 0.27 & 0.79 &  373 \\
\emph{Drosophila melanogaster} 
&       0.40 & 0.64 & 0.08 &{\bf 0.22}& 0.88 & 0.24&{\bf 0.23}& 0.82 & 0.35 & 0.74 &  78 \\
\emph{Caenorhabditis elegans} 
&{\bf  0.30}& 0.78 & 0.18 &{\bf 0.20}& 0.89 & 0.37 &{\bf 0.23}& 0.82& 0.34 & 0.75 &  114 \\
\emph{Arabidopsis thaliana} 
&       0.39& 0.62 & 0.11 &{\bf 0.29}& 0.78 & 0.19 & 0.43& 0.60 & 0.15 & 0.75 &  131 \\
Epstein-Barr virus 
&       0.31& 0.78 & 0.00 &{\bf 0.16}& 0.96 & 0.22 & 0.16& 0.87 & 0.48 & 0.81 &  23 \\
\end{tabular}
\begin{center}
{\small
Average rank-scores that indicate significantly increased according to both measures discussed in the Materials and Methods section ($p$-value $<10^{-3}$) are given in bold.}
\end{center}
\label{table1}
\end{table*}

\begin{figure}
\begin{center}
\centerline{\includegraphics[width=0.5\textwidth]{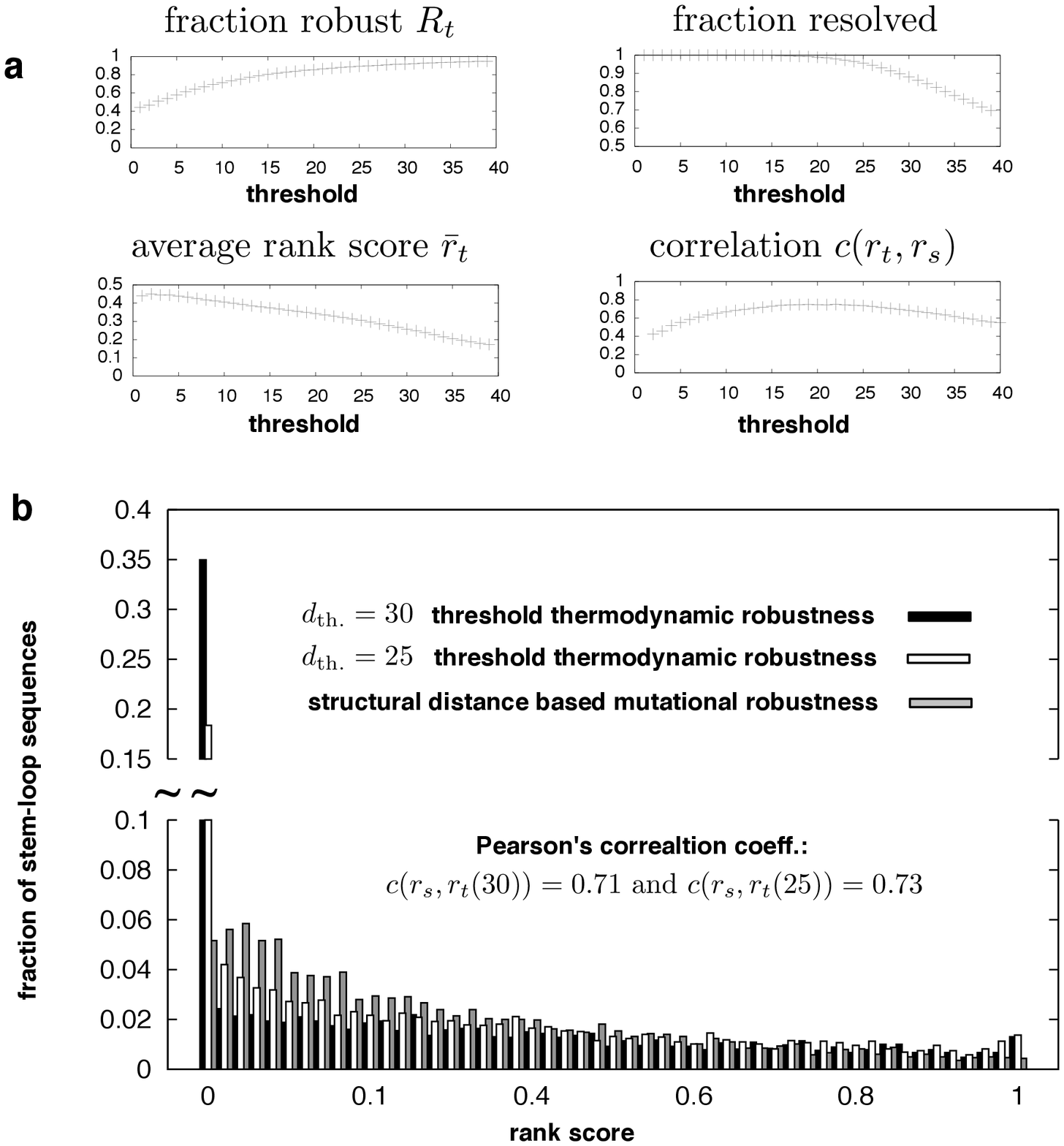}}
\caption{
To quantify the extent of thermodynamic robustness present in miRNA precursor sequences we examined the rank statistics of the threshold thermodynamic robustness $\eta_t(d_{\rm th.})$ which we defined as the cumulative frequency of structures in the thermodynamic ensemble that are equal to or less than a distance threshold $d_{\rm th.}$. {\bf a} For threshold values lager than $d_{\rm th.}\approx20$ the average rank of miRNA precursor sequences with respect to $\eta_t(d_{\rm th.})$, denoted by $\bar r_t(d_{\rm th.})$, becomes larger than $\bar r_s$, while remaining highly correlated with it. For thresholds $d_{\rm th.}>25$ the $\eta_t(d_{\rm th.})$ values for an increasingly larger fraction of random sample sequences becomes indistinguishable from the $\eta_t(d_{\rm th.})$ value of the original miRNA precursor sequence {\color{black} due to a lack of structures with $d>d_{th.}$ in our finite resolution sample of the equilibrium ensemble. The fraction of sequences which remain distinguishable are labeled \emph{fraction resolved}}. {\bf b} Among the majority of the random sample sequences that are resolved, however, the threshold robustness of large number of the miRNA precursor sequences becomes highly significant (black and white bars) compared to the mutation robustness as measured by $\eta_s$ (grey bars), whilst retaining a high correlation among $r_s$ and $r_t(d_{\rm th.})$.}\label{fig3}
\end{center}
\end{figure}

\section*{Results}
We assessed the environmental and mutational robustness of $3641$ unique miRNA precursor sequences and for each sequence compared them to a random sample of sequences with the same MFE structure. The idea of looking for signs of adaptation for increased robustness among miRNA precursor sequences by comparing the robustness of naturally occurring sequences to that of random sequences with the same secondary structure is conceptually similar to the approach used to support the argument that the genetic code has evolved to minimize mutational load \cite{digulio,haighurst,szathmarybook}. In the case of the genetic code the authors took the common genetic code, and, for each codon, calculated the change in polarity of the encoded amino acid caused by replacing each of the three nucleotides, one after the other. In order to determine whether the genetic code is adapted to minimize mutational load they proceeded by comparing the mean squared change caused by the replacement of a single nucleotide in the common genetic code to $10000$ randomly generated codes with the same redundancies. They found that only two of the random codes were more conservative than the common code with respect to polarity distances between neighbouring amino acids.       

We undertook a similar program in the case of miRNA precursor sequences. Each miRNA gene encodes a short $\approx22$ nucleotide sequence that is partially complementary to the mRNA of proteins regulated by the particular miRNA gene. For the proper short sequence to be excised by the protein Dicer, and hence for the miRNA gene to be functional, a larger part of the miRNA sequence, called the miRNA precursor sequence, must fold into the proper secondary structure. In order to determine whether a miRNA precursor sequence is adapted to minimize the effects of mutational and/or environmental perturbations, i.e.\ to maximize mutational and/or environmental robustness, we compared the mutational and environmental robustness of each miRNA precursor sequence {\color{black}(robustness measures we used are defined below)} to the mutational and environmental robustness of a random sample of sequences with identical structural phenotype (i.e.\ identical MFE structure). 

To generate a random sample of sequences with given MFE structure we first used, starting from a random sequence, stochastic minimization of the free-energy of the target structure to find a sequence with the desired MFE structure. This method by itself, however, yields a \emph{biased} sample of sequences (see Fig.\ \ref{fig1}a and b) and must be supplemented by an additional randomization step (see Materials and Methods). To measure the mutational robustness of a given sequence we used the measures introduced by Borenstein and Ruppin \cite{ruppin}: (i) the structural distance based mutational robustness measure $\eta_s$  of an RNA sequence of length $L$ is defined by $\eta_s= 1/(3L) \sum_{i=0}^{3L} (L-d_i)/L $, where $d_i$ is the base-pair distance between the secondary structure of mutant $i$ and the native sequence ({\color{black} given by the number of base pairs present in one structure, but not the other}), and the sum goes over all $3L$ single mutant neighbours and (ii) the more stringent measure $\eta_n$ is simply defined as the fraction of neutral single mutant neighbours, i.e.\  those that have identical MFE structure to the original sequence. In order to quantify the level of excess mutational robustness among miRNA precursor sequences we counted, for each miRNA precursor sequence, the number of sample sequences that have higher mutational robustness according to a given measure (see Materials and Methods) and used this to calculate the rank scores $r_s$ and $r_n$, defined as the fraction of sample sequences with identical or higher robustness according to $\eta_s$ and $\eta_n$, respectively. To facilitate an overview of the extent of excess mutational robustness we also calculated the average of the  rank scores over all miRNA precursor sequences $\bar r_s$ and $\bar r_n$ as well as the fraction of miRNA precursor sequences with higher than average robustness (i.e.\ rank-scores $<0.5$) $R_s$ and $R_n$ and the fraction of sequences with statistically significant increased robustness (i.e.\ rank-scores $<0.05$, see Materials and Methods) $S_s$ and $S_n$, respectively, according to a give measure. The statistical significance of both rank scores for individual miRNA precursor sequences as well as that of the finding a given fraction of robust sequences among a group of sequences was determined as detailed in the Materials and Methods section. 

Reexamining the mutational robustness of miRNA precursor sequences in comparison to an \emph{unbiased} sample of sequences with identical MFE structure we found that miRNA precursor sequences -- in contrast to the results of Borenstein and Ruppin -- do not have significantly more neutral single mutant neighbours than sample sequences, but do show a statistically significant increase in robustness measured according to $\eta_s$ (see Fig.\ \ref{fig1}b and Table \ref{table1}). In other words, native miRNA precursor sequences have on average the same number of single mutant neighbours with MFE structures identical to their own, as random sample sequences with the same structure. The MFE structure of those single mutant neighbours that are not identical to their own are, on the other hand, significantly more similar than in the case of sample sequences.         

The presence of excess mutational robustness is, by itself, insufficient to determine whether mutational robustness has evolved as a result of direct selection or in congruence with selection for environmental robustness. As established previously \cite{ruppin} there is evidence for excess thermodynamic robustness, robustness to thermal fluctuations, as evidenced by a significantly lower than chance minimum folding energy among miRNA precursor sequences. Defining the environmental robustness measure $\eta_E$ simply as minus the minimum folding energy we also find $\bar r_E=0.278$, $R_E=0.796$, $S_E=0.220$ using unbiased sampling. The correlation between $r_s$ and $r_E$ across miRNA precursor sequences is, however, rather low with a Pearson's correlation coefficient of $c(r_s,r_E)=0.217$ and $c(r_n,r_E)=0.071$. The minimum folding energy is a somewhat crude measure of thermodynamic robustness and does not even reflect the excess mutational robustness according to the measure $\eta_s$. {\color{black} There is no good reason to assume that a low MFE in itself confers environmental robustness, as even for relatively high free energies a given sequence may none the less with high probability fold into structures sufficiently similar to the MFE structure to remain functional.} The large number of miRNA precursor sequences that exhibit excess mutational robustness as measured by the structural similarity based measure $\eta_s$ suggests that a strict adherence to the MFE structure is not necessary to retain functionality -- a sufficiently similar, but not necessarily identical, secondary structure is enough to guarantee the excision of the proper subsequence. {\color{black} This is further supported by the fact that folding free energy alone is not sufficient to discriminate miRNA precursors, as well as recent evidence that a diverse set of structural features are needed for successful cleavage of a miRNA precursor sequence \cite{ritchie}. } 

To construct an appropriate  measure of thermodynamic robustness that also reflects this observation we would need to know the extent of similarity that is required to retain functionality -- indeed we would require detailed knowledge of the interaction between the RNA substrate and the enzyme Dicer to establish an appropriate measure of structure similarity. As such information is not at present available we chose to use the most simple and widely employed structure similarity measure, the base-pair distance used above. In order to determine the extent of similarity required to retain functionality we defined the threshold thermodynamic robustness measure $\eta_t(d_{\rm th.})$ as a function of the threshold distance $d_{\rm th.}$, by equating it with the probability in the equilibrium thermodynamic ensemble of structures  that have base-pair distances equal to or less than a threshold $d_{\rm th.}$ with respect to the MFE structure, i.e.
\begin{equation}
\eta_t(d_{\rm th.}) = \sum_{i \in \Omega} H(d_{\rm th.}-d_i)\ \frac{{\rm e}^{-E_i/kT}}{Z},
\end{equation}
where the sum goes over the set of all possible structures $\Omega$, $d_i$ denotes the base-pair distance of structure $i$ to the MFE structure,  $Z= \sum_{i \in \Omega}{\rm e}^{-E_i/kT} $ is the partition sum and $H(x)$ is the unit step function, i.e.\ $H(x)=0$ if $x<0$ and $H(x)=1$ if $x \geq 0$.
  
Examining the thermodynamic robustness of miRNA precursor sequences in comparison to an unbiased sample of sequences with identical MFE structure we found that  miRNA precursor sequences have significantly more structures in their equilibrium thermodynamic ensemble that are similar to the MFE structure than sample sequences (see Fig.\ \ref{fig2}a,b and Table \ref{table1}). In other words miRNA precursor sequences tend to adapt more similar structures as a result of thermal fluctuations than random sample sequences with the same structure. Calculating the average rank score $\bar r_t(d_{\rm th.})$ and the fraction of robust $R_t(d_{\rm th.})$ and significantly robust $S_t(d_{\rm th.})$ miRNA precursor sequences, with respect to the measure $\eta_t(d_{\rm th.})$ (Fig.  \ref{fig3}a) and examining the distribution of structures as a function of the base-pair distance for individual miRNA precursor sequences (see e.g.\ Fig.\ \ref{fig2}a) indicates that above a  threshold distance $d_{\rm th.}\approx20$ the measures start to saturate, yielding an estimate of the required similarity to retain function. The correlation between the rank-score of miRNA precursor sequences according to the distance similarity based mutational robustness measure and the threshold thermodynamics measure is high for all threshold values.  This is the direct result of the high degree of similarity between the distribution of structures in the thermodynamic ensemble and the mutational neighborhood (Fig.\ \ref{fig2}). The average rank score $\bar r_t(d_{\rm th.})$ and the fraction of robust $R_t(d_{\rm th.})$ and significantly robust $S_t(d_{\rm th.})$ miRNA precursor sequences  with respect to the threshold thermodynamic robustness measure indicate a markedly larger extent of excess robustness than their counterparts for mutation robustness, i.e.\  $\bar r_s$, $R_s$  and $S_s$ (see Fig.\ \ref{fig3}a,b and Table \ref{table1}) above $d_{\rm th.}>20$.  
         
\section*{Discussion}
The results presented above demonstrate the correlated presence of excess environmental (thermodynamic) and genetic (mutational) robustness among miRNA precursor sequences as measured according to, respectively, $\eta_s$ and $\eta_t(d_{\rm th.})$. A rather general causality between environmental and genetic robustness in the context of RNA secondary structure has been suggested by Ancel and Fontana \cite{ancelfontana}, who studied the dynamics of an \emph{in silico} population of RNA sequences evolving towards a predefined target shape. They found that a correlation exists between the set of shapes in the plastic repertoire of a sequence and the set of dominant (minimum free energy) shapes in its genetic neighborhood. They argue that this statistical property of the RNA genotype-phenotype map, which they call plastogenetic congruence, traps populations in regions where most genetic variation is phenotypically neutral. In other words RNA sequences explore a similar repertoire of suboptimal structures as a result of perturbations due to mutations and perturbations resulting from thermal fluctuations, and selection for a given target structure favours sequences with higher robustness to perturbations of both type.     

Since, in contrast to genetic robustness, environmental robustness does not require high values of $uN_e$, as it is a property of the sequence and not its mutational neighborhood, we contend that the observed bias in mutational robustness is in fact the result of the \emph{congruent} evolution of environmental and genetic robustness.  

The correlation between the response to heritable (mutational) and nonheritable (thermodynamic) perturbation, and hence the congruent evolution of genetic and environmental robustness may extend to other systems with genotype-phenotype maps different from RNA secondary structure. In particular, Xia and Levitt \cite{xia}, have found compelling evidence of the correlated evolution of increased thermodynamic stability and the number of neutral neighbours in lattice protein models. Understanding the relationship between sequence, structure, and function is, and will remain to be in the foreseeable future, a central theme in both molecular and evolutionary biology. A comprehensive view of how  the relationship between sequence, structure, and function is shaped during the course of evolution must take into consideration both the potential correlations that arise from the physics of the structure-sequence relationship as well as the relevant population genetic conditions in the context of which it takes on the role of a genotype-phenotype map.  {\color{black} In the context of computational miRNA gene discovery our thermodynamic robustness measure potentially offers an improved structural feature that may perform better than the free energy score of the hairpin or its ensemble diversity (which have proved uninformative \cite{freyhult}).}

\begin{acknowledgments}
This work was partially supported by the Hungarian Science Foundation (K60665).
\end{acknowledgments}


\begin{thebibliography}{100}

\bibitem[Ancel and Fontana 2000]{ancelfontana}
Ancel LW, Fontana W (2000)
Plasticity, Evolvability, and Modularity in RNA
\emph{J.\ Exp.\ Zool.\ } 288: 242-283.\\
(DOI:10.1002/1097-010X(20001015)288:3<242::AID-JEZ5>3.0.CO;2-O)

\bibitem[Azevado et al.\ 2007]{azevedoNAT}
Azevedo \emph{et al.\ } (2007)
Sexual reproduction selects for robustness and negative epistasis in artificial gene networks. 
\emph{Nature} 440: 87-90.\\
(DOI:10.1038/nature04488)

\bibitem[Bartel 2004]{bartelCELL}
Bartel DP (2004) 
MicroRNAs: genomics, biogenesis, mechanism, and function.  
\emph{Cell} 116: 281--297.\\
(DOI:doi:10.1016/S0092-8674(04)00045-5)

{\color{black}
\bibitem[Bloom et al.\ 2007]{bloom}
Bloom JD, Lu Z, Chen D, Raval A, Venturelli OS, Arnold FH (2007)
Evolution favors protein mutational robustness in sufficiently large populations.
\emph{BMC Biol.\ } 5: 29.\\
(DOI:10.1186/1741-7007-5-29)
}
\bibitem[Borenstein and Ruppin 2006]{ruppin}
Borenstein E, Ruppin E (2006)  
Direct evolution of genetic robustness in microRNA
\emph{Proc.\ Natl.\ Acad.\ Sci.\ U.S.A.}
103: 6593-6598.\\
(DOI:10.1073/pnas.0510600103)

\bibitem[Ciliberti et al.\ 2007a]{wagnerPLOS}
Ciliberti S, Martin OC, Wagner A (2007) 
Robustness can evolve gradually in complex regulatory networks with varying topology. 
\emph{PLoS Computational Biology} 3: e15.\\
(DOI:10.1371/journal.pcbi.0030015)

\bibitem[Ciliberti et al.\ 2007b]{wagnerPNAS}
Ciliberti S, Martin OC, Wagner A (2007) 
Innovation and robustness in complex regulatory gene networks. 
\emph{Proc.\ Natl.\ Acad.\ Sci.\ U.S.A.} 104: 13591-13596.\\
(DOI:10.1073/pnas.0705396104)

\bibitem[Crombach and Hogeweg 2008]{crombachPLOS}
Crombach A,  Hogeweg P (2008)
Evolution of Evolvability in Gene Regulatory Networks 
\emph{PLoS Comput.\ Biol.\ } 4: e1000112.\\ 
(DOI:10.1371/journal.pcbi.1000112)

\bibitem[Di Giulio 1987]{digulio}
Di Giulio M (1987) 
The extension reached by the minimization of the polarity distances during the evolution of the genetic code. 
\emph{J.\ Mol.\ Evol.\ } 29: 288-293.\\
(DOI:10.1007/BF02103616) 

\bibitem[Fisher 1928]{fisher}
Fisher RA (1928) The possible modifications of the response of
the wild type to recurrent mutations. \emph{Am.\ Nat.\ } 62:115-126.\\
(DOI:10.1086/280193)

\bibitem[Forester et al.\ 2006]{adamiJTB}
Forster R, Adami C, Wilke CO (2006)
Selection for mutational robustness in finite populations
\emph{J.\ Theor.\ Biol.\ } 243: 181-190.\\
(DOI:10.1016/j.jtbi.2006.06.020) 

{\color{black}
\bibitem[Freyhult et al. 2005]{freyhult}
Freyhult E, Gardner PP, Moulton V (2005)
A comparison of RNA folding measures.
\emph{BMC Bioinformatics} 6: 241.\\
(DOI:10.1186/1471-2105-6-241)
}

\bibitem[Griffiths-Jones et al. 2006]{mirbase} 
Griffiths-Jones S, Grocock RJ, van Dongen S, Bateman A, Enright AJ (2006)
miRBase: microRNA sequences, targets and gene nomenclature.
\emph{Nucleic Acids Research} 34: D140.\\
(DOI:10.1093/nar/gkj112) 

\bibitem[Haigh and Hurst 1991]{haighurst}
Haig D, Hurst LD (1991)
A quantitative measure of error minimization in the genetic code.  
\emph{J.\ Mol.\ Evol.\ } 33: 412-417.\\
(DOI:10.1007/BF02103132)

\bibitem[Haldane 1930]{haldane}
Haldane, JBS (1930) 
A note on Fisher's theory of dominance.
\emph{Am.\ Nat.\ } 64:87-90.\\
(DOI:10.1086/280299)

\bibitem[Hillenmeyer et al.\ 2008]{hillenmeyer}
Hillenmeyer ME \emph{et al.\ } (2008) 
The Chemical Genomic Portrait of Yeast: Uncovering a Phenotype for All Genes 
\emph{Science} 320: 362.\\
(DOI:10.1126/science.1150021)

\bibitem[Hofbacker et al.\ 1994]{vienna}
Hofacker IL,  \emph{et al.} (1994) 
Fast Folding and Comparison of RNA Secondary Structures. 
\emph{Monatshefte f\"ur Chemie} 125: 167-188.\\
(DOI:doi:10.1007/BF00818163)

\bibitem[Kacser and Burns 1981]{kacserburns}
Kacser H, Burns JA (1981) 
The molecular basis of dominance.  
\emph{Genetics} 97:6639-6666.\\
(DOI:n.a.\ )

\bibitem[Kimura 1966]{kimura}
Kimura M,Maruyama T (1966) 
The mutational load with epistatic gene interactions in fitness. 
\emph{Genetics} 54: 1337-1351.\\
(DOI:n.a.\ )

\bibitem[Krakauer and Plotkin 2002]{krakauerPNAS}
Krakauer DC, Plotkin JB (2002) 
Redundancy, antiredundancy, and the robustness 
of genomes. 
\emph{Proc.\ Natl.\ Acad.\ Sci.\ USA} 99: 1405. \\
(DOI:10.1073/pnas.032668599)

\bibitem[Lagos-Quintana et al.\ 2001]{lagosQSCI}
Lagos-Quintana M, Rauhut R, Lendeckel W, Tuschl T (2001)
Identification of novel genes coding for small expressed RNAs. 
\emph{Science}  294:853-858.\\
(DOI:10.1126/science.1064921)

\bibitem[Lau et al.\ 2001]{lauSCI}
Lau NC, Lim LP, Weinstein EG, Bartel DP (2001)
An abundant class of tiny RNAs with probable regulatory roles in Caenorhabditis elegans.
\emph{Science} 294:858-862.\\
(DOI:10.1126/science.1065062)

\bibitem[Lee and Ambros 2001]{leeSCI}
Lee RC, Ambros V
(2001) An extensive class of small RNAs in Caenorhabditis elegans.
\emph{Science} 294:862-864.\\
(DOI:10.1126/science.1065329)

\bibitem[Lynch and Conery 2003]{lynch}
Lynch M, Conery JS (2003) The Origins of Genome Complexity
\emph{Science} 302: 1401 - 1404.\\
(DOI:10.1126/science.1089370)

\bibitem[Mayo and  B\"urger 1997]{burger}
Mayo O, B\"urger R (1997) Evolution of dominance: a theory
whose time has passed? \emph{Biol.\ Rev.\ } 72: 97-110.\\
(DOI:10.1111/j.1469-185X.1997.tb00011.x) 

\bibitem[Monetville et al. 2005]{montville}
Montville R \emph{et al.\ }(2005) Evolution of Mutational Robustness in an RNA Virus
\emph{PLOS Biology} 3:e381\\
(DOI:10.1371/journal.pbio.0030381)

\bibitem[van Nimwegen et al.\ 1999]{nimwegen}
van Nimwegen E, Crutchfield P, Huynen M  (1999)
Neutral evolution of mutational robustness
\emph{Proc.\ Natl.\ Acad.\ Sci.\ U.S.A.}
96: 9716-9720.\\
(DOI:n.a.\ )

{\color{black}
\bibitem[Ritchie et al.\ 2007]{ritchie}
Ritchie W, Legendre M, Gautheret D (2007)
RNA stem-loops: to be or not to be cleaved by RNAse III. 
\emph{RNA} 13: 457-462. \\
(DOI:10.1261/rna.366507)
}

{\color{black}
\bibitem[S\'anjuan et al.\ 2007]{sanjuan}
Sanju\'an R, Cuevas JM, Furi\'o V, Holmes EC, Moya A (2007)
Selection for robustness in mutagenized RNA viruses.
\emph{PLoS Genet.} 3: e93.\\
(DOI:10.1371/journal.pgen.0030093.eor)
}

\bibitem[Shu et al.\ 2007]{Shu}
Shu W {\emph et al.} (2007) 
In silico genetic robustness analysis of microRNA secondary structures: potential evidence of congruent evolution in microRNA. \emph{BMC Evolutionary Biology} 7: 223. \\ 
(DOI:10.1186/1471-2148-7-223)

\bibitem[Siegal and Bergman 2002]{siegalPNAS}
Siegal LS, Bergman A (2002) Waddington's canalization revisited: Developmental stability and evolution  
\emph{Proc.\ Natl.\ Acad.\ Sci.\ USA} 99: 10528.\\
(DOI:10.1073/pnas.102303999)

\bibitem[Szathm\'ary and Smith 1997]{szathmarybook}
Szathm\'ary E, Smith JM (1997) 
\emph{The major transitions in evolution} (Oxford University Press, Oxford, UK)\\
(ISBN:019850294X)

\bibitem[Sz\"oll\H{o}si and Der\'enyi 2008]{szollosiMBS}
Sz\"oll\H{o}si GJ, Der\'enyi I (2008)
The effect of recombination on the neutral evolution of genetic robustness.
\emph{Math.\ Biosci.\ } 214: 58-62.\\
(DOI:10.1016/j.mbs.2008.03.010)

\bibitem[de Visser et al.\ 2003]{devisser}
de Visser JAG \emph{et al.\ } (2003) Perspective: Evolution and Detection of Genetic Robustness 
\emph{Evolution} 57: 1959-1972\\
(DOI: 10.1554/02-750R)

\bibitem[Wright 1934]{wright}
Wright S (1934) Physiological and evolutionary theories of dominance. 
\emph{Am.\ Nat.\ } 68:25-53.\\
(DOI:10.1086/280521)

\bibitem[Waddington 1957]{waddington}
Waddington, CH, Kacser H (1957) \emph{The Strategy of the Genes: A Discussion of Some Aspects of Theoretical Biology} (MacMillan, New York USA). \\
(ISBN:n.a.\ )

{\color{black}
\bibitem[Wagner and Stadler 1999]{wagnerJexp}
Wagner A, Stadler PF (1999)
Viral RNA and evolved mutational robustness.
\emph{J Exp Zool.} 285:119-27\\
(DOI:10.1002/(SICI)1097-010X(19990815)285:2<119::AID-JEZ4>3.0.CO;2-D )
}

{\color{black}
\bibitem[Wilke and Adami 2003]{wilkeadami}
Wilke CO, Adami C (2003)
Evolution of mutational robustness.
\emph{Mutat. Res.\ } 522: 3-11\\
(DOI:10.1016/S0027-5107(02)00307-X)
}

\bibitem[Xia and Levitt 2002]{xia}
Xia Y, Levitt M (2002)
Roles of mutation and recombination in the evolution of protein thermodynamics
\emph{Proc.\ Natl.\ Acad.\ Sci.\ U.S.A.} 99:10382-10387. \\
(DOI: 10.1073/pnas.162097799)

\end{thebibliography}
\end{document}